\documentclass[a4paper,11pt]{article}
\usepackage{jheppub}
\usepackage[T1]{fontenc} 

\usepackage{url}
\usepackage{graphicx}

\usepackage{latexsym}
\usepackage{amssymb}
\usepackage{amsfonts}

\usepackage{amsmath}
\usepackage[caption=false]{subfig}

\usepackage{breqn}

\newcommand{\bea}{\begin{eqnarray}}
\newcommand{\eea}{\end{eqnarray}}
\newcommand{\be}{\begin{equation}}
\newcommand{\ee}{\end{equation}}

\newcommand{\dd}{{\rm d}}
\newcommand{\nn}{\nonumber}

\newcommand{\ep}{\epsilon}

\newcommand{\cmnt}[1]{}

\allowdisplaybreaks[4]
 
\title{Evaluating `elliptic' master integrals at special kinematic values: using
differential equations and their solutions via expansions near singular
points}

\author[a]{Roman N. \ Lee}
\author[b,d]{Alexander V.\ Smirnov}
\author[c]{Vladimir A.\ Smirnov}
\affiliation[a]{
Budker Institute of Nuclear Physics, 630090 \\ Novosibirsk, Russia}
\affiliation[b]{Research Computing Center, Moscow State University, 
119992 Moscow, Russia}
\affiliation[c]{Skobeltsyn Institute of Nuclear Physics of Moscow State 
University, 119992 Moscow, Russia}
\affiliation[d]{Institut f\"{u}r Theoretische Teilchenphysik, KIT, 76128 
Karlsruhe, Germany}
\emailAdd {roman.n.lee@gmail.com}
\emailAdd {asmirnov80@gmail.com}
\emailAdd {smirnov@theory.sinp.msu.ru}

\abstract{
 This is a sequel of our previous paper where we described an algorithm to find a solution 
of differential equations for master integrals
in the form of an $\epsilon$-expansion series with numerical coefficients. 
The algorithm is based on using generalized power series expansions near singular points 
of the differential system, solving difference equations for the corresponding coefficients in these expansions
and using matching to connect series expansions at two neighboring points.
Here we use our algorithm and the corresponding code for our example
of four-loop generalized sunset diagrams with three massive and two massless propagators,
in order to obtain new analytical results. We analytically evaluate the master integrals at threshold,
$p^2=9 m^2$, in an expansion in $\ep$ up to $\ep^1$. With the help of our code, we obtain
numerical results for the threshold master integrals in an $\ep$-expansion
with the accuracy of 6000 digits and then use the {\tt PSLQ} algorithm to arrive
at analytical values. Our basis of constants is build from bases of multiple polylogarithm values 
at sixth roots of unity.
}
\keywords{
Multiloop Feynman integrals, multiple polylogarithms, sixth roots of unity}

\begin{document}

\maketitle
\flushbottom

\section{Introduction}

Analytical results for Feynman integrals can, typically, be expressed in terms of harmonic 
polylogarithms~\cite{Remiddi:1999ew} or multiple polylogarithms~\cite{Goncharov:1998kja}
which are very well mathematically studied special functions\cmnt{ introduced by physicists}.
For harmonic polylogarithms, one can apply the package {\tt HPL} \cite{Maitre:2005uu} 
which encodes various analytical properties and provides the possibility of numerical evaluation with a desirable precision. 
For multiple polylogarithms, one can either use the computer code \cite{GiNaC_code} based on the {\tt GiNaC} 
library \cite{GiNaC} to obtain high-precision numerical values or construct a code based on their
algebraic properties, for example, to reveal their behavior near singular points.
Anyway, with a result in terms of these functions at hand, one can evaluate 
a Feynman integral at regular points and obtain expansions at singular points.
 
The possibility to arrive at a result written in terms of these functions exists if,
within the method of differential equations \cite{Kotikov:1990kg,Kotikov:1991pm,Remiddi:1997ny,Gehrmann:1999as,Gehrmann:2000zt,Gehrmann:2001ck},
one succeeds to turn to a so-called  {\em canonical} basis~\cite{Henn:2013pwa} of 
the master integrals\footnote{
There are various codes to arrive at a canonical form (or, $\epsilon$-form~\cite{Lee:2014ioa}) -- see
\cite{Lee:2014ioa,Gituliar:2016vfa,Gituliar:2017vzm,Prausa:2017ltv,Meyer:2016slj,Meyer:2017joq}.}.
It is well known that the $\epsilon$-form of DE for a given set of the master integrals is not 
always possible\footnote{Recently, a strict criterion of the existence of an $\epsilon$-form was presented in Ref. \cite{Lee:2017oca}.}
The simplest counter example is given by the two-loop propagator sunset diagram with three identical masses. In many irreducible cases, the lowest nontrivial term of $\epsilon$-expansion is expressed in terms of elliptic integrals, and in what follows we will also refer to these cases as `elliptic' in no relation with the functional form of the coefficients of $\ep$ expansion.
 
In situations without canonical bases, one can hope that the number of elliptic master integrals
is small and try to obtain, in these cases, two and three-fold parametric representations suitable 
for numerical evaluation, and in all other cases to proceed with canonical subbases -- see
examples of such an approach in Refs.~\cite{Aglietti:2007as,Bonciani:2016qxi,Primo:2016ebd,Primo:2017ipr}.
On the other hand, it is quite natural to try to introduce new functions which would enable us to present
results, in elliptic cases, in an analytical form. Multiple suggestions to introduce elliptic generalizations of multiple polylogarithms can be found in
Refs.~\cite{Adams:2016xah,Adams:2017ejb,Adams:2018yfj,Remiddi:2017har,Ablinger:2017bjx,Hidding:2017jkk,Broedel:2017kkb,Broedel:2017siw,Broedel:2018iwv}.
However, these new functions do not have the same status as harmonic and multiple polylogarithms, as far as a detailed description of their properties and the possibility to evaluate them numerically are concerned.
Moreover, the examples of successful treatment of $\ep$-expansion in elliptic cases are, at most at the two-loop level.
Anyway, we are very far, even in lower loops orders, from obtaining a complete description of
a class of functions which can appear in results for Feynman integrals.
 
We advocated~\cite{Lee:2017qql} an alternative way to solve differential equations
which can be used
also in `elliptic' situations and illustrated it through a four-loop example.
We considered multiloop Feynman integrals depending on one variable, i.e. with two scales 
where the variable is introduced as the ratio of these scales.
We described an algorithm to find a solution of a given differential system 
in the form of an $\epsilon$-expansion series with numerical coefficients. 
It is based on using generalized power series expansions near singular points 
of the differential system, solving difference equations for the corresponding coefficients in these expansions
and using matching to connect series expansions at two neighbouring points.
We provided a computer code where this algorithm is implemented for a simple example 
of a family of four-loop Feynman integrals where the $\epsilon$-form is impossible.
Using this code it is possible to evaluate master integrals at a given point
as well as expansions at singular points with
a required precision in an $\epsilon$-expansion with a required number of terms.

Our present paper is a sequel of Ref.~\cite{Lee:2017qql}. The goal here is to
apply our algorithm which is numerical in its character and the corresponding code in our example,
i.e. four-loop generalized sunset diagrams with three massive and two massless propagators,
in order to
obtain new analytical results. We analytically evaluate the master integrals at threshold,
$p^2=9 m^2$, in an expansion in $\ep$ up to $\ep^1$.
Remember that the $\ep$-form of the corresponding equations is impossible so that we cannot use
the well-known procedure of using a solution at a general point in terms of well-established
special functions and then turn to results at this singular point.
However, although solutions to the differential equations look too complicated,
the values of the master integrals at some special points can be conventional polylogarithmic constants. We will see that this is indeed true for the $\ep$-expansion of the integrals of our family at the threshold.
We use our code in order to construct a linear operator (a matrix) which renders the boundary conditions in one, suitable chosen, singular point to the coefficients of asymptotic expansion at the other point,  $p^2=9 m^2$ in our case. From these coefficients we extract the values of the integrals at $p^2=9 m^2$.

After reminding the main points of our setup in Section~2, we explain in Section~3 how
we obtain high-precision values, up to 6000 digits, for the threshold integrals and
then succeed in finding a relevant basis of constants in order to use the {\tt PSLQ} algorithm \cite{PSLQ}.
It turns out that the relevant bases of constants can be constructed starting from the bases of 
multiple polylogarithm values at sixth roots of unity,
i.e. of the form $G(a_1,\ldots,a_w;1)$
where the indices $a_i$ are equal to zero or a sixth root of unity, with $a_1\neq 1$,
which were constructed in Ref.~\cite{Henn:2015sem} up to weight six.
We discuss various perspectives in Conclusion.

\section{Setup}
   
Differential equations for master integrals have the form 
\begin{equation}
\partial_{x}\boldsymbol{J}=M\left(x,\epsilon\right)\boldsymbol{J}\,,
\label{DE}
\end{equation}
where $x$ is a dimensionless ratio of two scales 
for a family of dimensionally regularized Feynman integrals depending on two scales,
$\boldsymbol{J}$ is a column-vector of $N$ functions, and $M$ is an $N\times N$ matrix with 
elements which are  rational functions of $x$ and $\epsilon$. 

The general solution of this linear system has the form
\begin{equation}
\boldsymbol{J}\left(x\right)=U\left(x\right)\boldsymbol{J}_0\,,
\label{DEsolution}
\end{equation}
where $\boldsymbol{J}_0$ is a column of constants, and $U$ is an evolution operator represented in terms of a path-ordered exponential  
\begin{equation}
U\left(x\right)=P\exp\left[\int dxM\left(x\right)\right]\,.
\label{evolution}
\end{equation}
We want to expand this operator in the vicinity of each singular point.
Without loss of generality, let us consider the expansion near  $x=0$.
It has the form
\begin{equation}
U\left(x\right)=\sum_{\lambda\in S}x^{\lambda}\sum_{n=0}^{\infty}\sum_{k=0}^{K_{\lambda}}\frac1{k!}C\left(n+\lambda,k\right)x^{n}\ln^{k}x\,,
\label{evolution_expansion}
\end{equation}
where $S$ is a finite set of powers, 
$K_\lambda\geqslant0$ is an 
integer number corresponding to the the maximal power of the logarithm. 

We assume that all the singular points of the differential system are regular so that
we can reduce the differential system to a local Fuchsian form in any singular point.  
Therefore, we can reduce it 
at $x=0$ to normalized Fuchsian form \cite{Lee:2017oca} by means of rational transformations. 
Let us assume that the system is in a global normalized Fuchsian form, i.e.,
\begin{equation}
M\left(x\right) = \frac{A_0}{x}+\sum_{k=1}^{s}\frac{A_k}{x-x_k} 
\label{NFF}
\end{equation}
and for any $k=0,\ldots,s$ the matrix $A_k$ is free of {\em resonances}, i.e. the difference of any two of its 
distinct eigenvalues is not integer. 
In particular, the `elliptic' cases, as a rule, can easily be reduced to a global normalized Fuchsian form. 

As it is shown in Ref.~\cite{Lee:2017qql}, $S$ and $K_\lambda$ can be determined and
the difference equations for the coefficients $C\left(n+\lambda,k\right)$ in 
(\ref{evolution_expansion}) can be solved algorithmically. In fact, the idea to use series expansions at singular points
and difference equations for the corresponding coefficients is very well known in mathematics.
In high-energy physics, this strategy when evaluating Feynman integrals can be found, 
for example, in Refs.~\cite{Pozzorini:2005ff,Aglietti:2007as,Kniehl:2017ikj,Mueller:2015lrx,Melnikov:2016qoc}.
Let us emphasize that our algorithm, with the current assumptions, provides solutions
with no more than a linear growth of computational 
complexity with respect to a required number of terms. This is very important for a subsequent 
matching procedure which enables one to connect series expansions at two neighbouring points and thereby
to obtain the possibility to evaluate Feynman integrals at any given point -- see details in Ref.~\cite{Lee:2017qql}.

\section{Four-loop generalized sunset diagram at threshold}

\begin{figure}[h] 
\begin{center}
\includegraphics[width=0.25\textwidth]{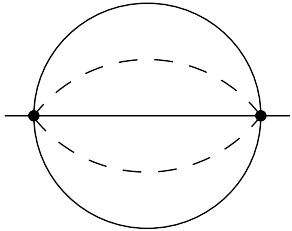}
\caption{\label{fig::ss00mmm}
The generalized sunset graph with two massless and three massive lines with the same mass.}
\end{center}
\end{figure}

As in Ref.~\cite{Lee:2017qql} let us consider the example of the following 
family of four-loop Feynman integrals:
\begin{eqnarray}
F_{a_1,\ldots,a_{14}}&=&\int\ldots\int\frac{
\dd^D k_1\ldots\dd^D k_4\;(k_1 \cdot p)^{a_6} (k_2\cdot p)^{a_7} (k_3\cdot p)^{a_8} (k_4\cdot p)^{a_9} }
{(-k_1^2)^{a_1}(-k_2^2)^{a_2}(m^2-k_3^2)^{a_3}(m^2-k_4^2)^{a_4}}
\nn  \\ \hspace*{-80mm} && \times
\frac{(k_1\cdot k_2)^{a_{10}} 
(k_1\cdot k_3)^{a_{11}} (k_1\cdot k_4)^{a_{12}} (k_2\cdot k_3)^{a_{13}} (k_2\cdot k_4)^{a_{14}}}
{(m^2-(\sum k_i+p)^2)^{a_5}}\;,
\label{fam18ind}
\end{eqnarray}
where $p$ is the external momentum and $m$ is the mass of three lines. They correspond to the generalized
sunset graph shown in Fig.~\ref{fig::ss00mmm}. We introduce $x = p^2/m^2$. 
  
There are four master integrals in this family.
As the primary master integrals we choose the following basis\footnote{
In our paper, we
use  {\tt FIRE}~\cite{Smirnov:2008iw,Smirnov:2013dia,Smirnov:2014hma} 
in combination with {\tt LiteRed}~\cite{Lee:2012cn,Lee:2013mka} to
solve integration by parts relations and reveal master integrals.}
\begin{equation}
\{F_{1,1,1,1,1,0,\ldots,0}, \, F_{1,1,2,1,1,0,\ldots,0}, \, F_{1,2,1,1,1,0,\ldots,0}, \,F_{1,2,1,1,2,0,\ldots,0}\} \,.
\label{MIs}
\nn
\end{equation}  
The singular points of the differential equations are $x_0 = 0, x_1 = 1, x_2 = 9$ and $x_3 \equiv x_{-1} = \infty.$
Using our algorithm we presented in Ref.~\cite{Lee:2017qql} the code {\tt DESS} to evaluate master integrals at a given point
as well as expansions at singular points with
a required precision keeping a required number of terms in an $\epsilon$-expansion.

Let us now formulate our current goal: to evaluate master integrals of family~(\ref{fam18ind})
considered at threshold, $p^2=9 m^2$, i.e. exactly at the singular point $x_2 = 9$. 
In fact, for such integrals defined with the same general formula ~(\ref{fam18ind}) we have now
three master integrals which can be chosen as
\begin{equation}
\{J_1=F_{1,1,1,1,1,0,\ldots,0}, \, J_2=F_{1,1,2,1,1,0,\ldots,0}, \, J_3=F_{1,2,1,1,1,0,\ldots,0}\} \,.
\label{MIs-th}
\end{equation}  
This can be done
with the code {\tt DESS}, where boundary conditions at the point $x_0 = 0$ were implemented.
This point is, however, not a neighbour of $x_2 = 9$ so that matching is used twice when transporting information
between $x_0$ and $x_2$. This results in the necessity to evaluate much more terms of the series expansions
at the three points  $\{x_0 , x_1 , x_2 \}$ in order to achieve a required precision which,
as we will see later, should be very high because of a big number of constants relevant to
results at $x_2$.

In fact, the choice of the point $x_0$ to impose boundary conditions encoded in {\tt DESS}
can be explained by the fact that, generally speaking, setting $p^2=0$ for propagator integrals
is equivalent to $p=0$ and resulting vacuum Feynman integrals turn out to have just less indices.
However, vacuum integrals can involve `more complicated' constants. To solve our current goal, we can
make a better choice to impose boundary conditions at the point $x_1$ for two reasons: this is now 
a neighbour of $x_2 = 9$ and the corresponding constants are multiple zeta values, logarithm of two and polylogarithms
of one half.
Indeed, master integrals at $x=x_1$ appeared in the calculations presented in Refs.~\cite{Marquard:2015qpa,Marquard:2016dcn}
where they were evaluated using a onefold Mellin-Barnes representation.
In particular, for $F_{1,1,1,1,1,0,\ldots,0}$ at $p^2=m^2=1$, the result is
\begin{eqnarray}
-\frac{1}{4 \ep^4} - \frac{7}{8\ep^3} - \left(\frac{17}{12} + \frac{\pi^2}{12}\right)\frac{1}{\ep^2}
+\left (\frac {\zeta (3)} {3} - \frac {835} {576} - \frac {7 \pi ^2}{24} \right)\frac{1}{\ep}
&& \nn \\  && \hspace*{-105mm}
+ \left(\frac{7 \zeta (3)}{6}-\frac{7379}{6912}-\frac{17 \pi ^2}{36}-\frac{3 \pi ^4}{20}\right)
\nn \\  && \hspace*{-105mm} 
+ \left(\frac{17 \zeta (3)}{9}+\frac{127 \pi ^2 \zeta (3)}{9}-\frac{289 \zeta
   (5)}{5}+\frac{6766055}{82944}-\frac{41539 \pi ^2}{1728}-\frac{21 \pi ^4}{40}\right) \ep
\nn \\  && \hspace*{-105mm} 
+ \left(   
-288 \pi ^2 \text{Li}_4\left(\frac{1}{2}\right)-\frac{355325 \zeta (3)}{432}+\frac{889 \pi ^2 \zeta
   (3)}{18}+\frac{2203 \zeta (3)^2}{9}-\frac{2023 \zeta (5)}{10}
\right.
\nn \\  && \hspace*{-105mm}
\left.
   -252 \pi ^2 \zeta (3) \log
   (2)+\frac{1449210865}{995328}-\frac{8822483 \pi ^2}{20736}-\frac{17 \pi ^4}{20}+\frac{3877 \pi ^6}{1890}
\right.
\nn \\  && \hspace*{-105mm}
\left.   
   -12
   \pi ^2 \log ^4(2)+12 \pi ^4 \log ^2(2)+424 \pi ^2 \log (2)
   \right) \ep^2   
\nn \\  && \hspace*{-105mm} 
+ \left(   
-5760 \zeta(-6,-1)-5760 \zeta(-6,1)-5760 \zeta(-5,2)
\right.
\nn \\  && \hspace*{-105mm}
\left.
-5760
   \zeta(-5,1,1)-5760 \zeta(5,-1,-1)+5760 s_6 \log (2)-10080
   \text{Li}_4\left(\frac{1}{2}\right) \zeta (3)
\right.
\nn \\  && \hspace*{-105mm}
\left.   
   +5184 \pi ^2 \text{Li}_5\left(\frac{1}{2}\right)-1008 \pi ^2
   \text{Li}_4\left(\frac{1}{2}\right)-16960 \text{Li}_4\left(\frac{1}{2}\right) 
\right.
\nn \\  && \hspace*{-105mm}
\left.     
  +5184 \pi^2 \text{Li}_4\left(\frac{1}{2}\right) \log (2)
   +\frac{312867 \zeta (7)}{14}-\frac{100204 \pi ^2 \zeta
   (5)}{15}-\frac{4913 \zeta (5)}{15}
 \right.
\nn \\  && \hspace*{-105mm}
\left.       
   +\frac{15421 \zeta (3)^2}{18}+\frac{908 \pi ^4 \zeta (3)}{15}
   +\frac{2159
   \pi ^2 \zeta (3)}{27}-\frac{77124781 \zeta (3)}{5184}
  \right.
\nn \\  && \hspace*{-105mm}
\left.   
-420 \zeta (3) \log ^4(2)
 +2688 \pi ^2 \zeta (3) \log^2(2)
   -7200 \zeta (3)^2 \log (2)
\right.
\nn \\  && \hspace*{-105mm}
\left.     
-882 \pi ^2 \zeta (3) \log (2)+\frac{3877 \pi ^6}{540}+\frac{195233 \pi^4}{1728}-\frac{1121725465 \pi ^2}{248832}
\right.
\nn \\  && \hspace*{-105mm}
\left.       
  +\frac{182188906799}{11943936}+\frac{864}{5} \pi ^2 \log ^5(2)-42
   \pi ^2 \log ^4(2)-\frac{2120 \log ^4(2)}{3}
\right.
\nn \\  && \hspace*{-105mm}
\left.         
   -144 \pi ^4 \log ^3(2)
   +42 \pi ^4 \log ^2(2)-\frac{9328}{3} \pi
   ^2 \log ^2(2)-\frac{14}{3} \pi ^6 \log (2)
\right.
\nn \\  && \hspace*{-105mm}
\left.          
   +7652 \pi ^2 \log (2) 
   \right) \ep^3   
+O(\ep^4)   \;,
\nn
\end{eqnarray}
where $\zeta(\ldots)$ are multiple zeta values.
Here the result is restricted to contributions of weight seven and
some efforts are needed to go further.

It turns out that the best way to impose boundary conditions is to choose
$x_3 = \infty$ because the corresponding expansion is nothing but
the large-momentum expansion \cite{Chetyrkin:1988zz,Chetyrkin:1988cu,Smirnov:1990rz,Smirnov:2002pj} where, 
for our integrals~(\ref{fam18ind}), any term
is a product of one-loop tadpoles and massless propagator integrals and 
can be evaluated via gamma functions at general $\ep$. This provides any required accuracy and any required number of terms
in $\ep$-expansions in the boundary conditions.
In the updated version of our code {\tt DESS}, we introduce the possibility to impose boundary conditions
at an arbitrary singular point. We added one more argument  {\tt ns} to the function 
\[
{\tt DESS[rdatas, x, f(x), oe, np, nt, ns]}
\] 
which means the number of a singular point and
this number is $1$ for $x_0$, $2$ for $x_1$, and $4$ for $x_3$. There is no sense to choose $x_2$ since this
point is most complicated from the calculational point of view.
We attach also two more auxiliary files: {\tt BoundaryConditions1} and {\tt BoundaryConditionsInf} where analytic results
for the boundary integrals are encoded.
As before, the code and the auxiliary data can be downloaded from 

\noindent \url{https://bitbucket.org/feynmanintegrals/dess}.
With the current version of {\tt DESS}, we have obtained numerical results for the threshold master integrals in an $\ep$-expansion up
to $\ep^2$ with the accuracy of 6000 digits for the corresponding coefficients. This took less than four hours on a desktop.
As we will see shortly, such a big accuracy is needed for an application of the {\tt PSLQ} algorithm.
  
The crucial point is a choice of a relevant basis of constants.
A first hint comes from the known results for the two-loop sunset diagram at threshold \cite{Berends:1997vk,Davydychev:1999ic}
where one can observe multiple polylogarithm values at sixth roots of unity 
and $\frac{\pi}{\sqrt{3}}$. Let us also take into account that, at least according to Refs.~\cite{Fleischer:1999mp,Davydychev:2000na,Kalmykov:2010xv},
it might be reasonable to include into the basis the constant $\sqrt{3}$ separately.
Therefore, we tried to use the bases connected with multiple polylogarithm values at sixth roots of unity
and constructed in Ref.~\cite{Henn:2015sem} up to weight six\footnote{Bases up to weight three were
constructed in Ref.~\cite{Broadhurst:1998rz} and up to weight four in Ref.~\cite{FrancescoMoriellosthesis}.}
and $\sqrt{3}$.
  
We consider bases of constants by including multiple polylogarithm values at sixth roots of unity up to weight six,
i.e. of the form $G(a_1,\ldots,a_w;1)$
where the indices $a_i$ are equal to zero or a sixth root of unity,  
i.e. taken from the seven-letters alphabet $\{0,r_1,r_3,-1,r_4,r_2,1 \}$ with
\be
r_{1,2}=\frac{1}{2}\left(1\pm\sqrt{3}\,{\rm i} \right)=\lambda^{\pm 1}\;, \;\;\;
r_{3,4}=\frac{1}{2}\left(-1\pm\sqrt{3}\,{\rm i} \right)=\lambda^{\pm 2}\;,\;\;\;
\lambda=e^{\pi {\rm i}/3}=r_1\;.
\label{letters}
\ee
and $a_1\neq 1$.

The multiple polylogarithms are defined as  
\be\label{eq:Mult_PolyLog_def}
 G(a_1,\ldots,a_w;z)=\,\int_0^z\frac{1}{t-a_1}\,G(a_2,\ldots,a_w;t)\, \dd t
\ee
with $a_i, z\in \mathbb{C}$ and $G(z)=1$. 
In the special case where  $a_i=0$ for all $i$, the corresponding integral is divergent 
and instead one defines
\be
G(0,\ldots,0;z) = \frac{1}{n!}\,\log^n z \;.
\ee
If $a_w\neq 0$ and $\rho\neq 0$, then 
$G(\rho a_1,\ldots,\rho a_w;\rho z) = G(a_1,\ldots,a_w;z) $ so that one can express
such MPL in terms of $G(\dots ;1)$.
The length $w$ of the index vector is called the weight.
One can consider separately the real and imaginary parts of the MPL
\be
G(a_1,\ldots,a_{w};1)= G_R(a_1,\ldots,a_{w})+ {\rm i}\, G_I(a_1,\ldots,a_{w})
\label{reim}
\ee
 
For example, the elements of weight one are chosen in Ref.~\cite{Henn:2015sem} as
\bea
G_R(-1)&=& \log (2)\;, \nn \\
G_R(r_4)&=& \frac{1}{2}\log(3)\;, \nn \\
G_I(r_2)&=& -\frac{\pi }{3}\nn\;.
\eea 
Let us denote by $B_R(w)$  ($B_I(w)$) the bases 
generated by $G_R(a_1,\ldots,a_{w})$ ($G_I(a_1,\ldots,a_{w})$). 
They include not only elements of the form $G_{R/I}(a_1,\ldots,a_{w})$ but also products of constants of lower weights.
The definitions of the bases can also be found
in auxiliary files supplied with Ref.~\cite{Henn:2015sem}. They can be downloaded from 
\url{http://theory.sinp.msu.ru/~smirnov/mpl6}.

As we shall see in our case in practice, when using the {\tt PSLQ} algorithm,
it is sufficient to use the bases
$B(w)=\{B_R(w),\sqrt{3} B_I(w)\}$ of weights $w=1,2,\ldots$. The element $\sqrt{3}$ does not contribute to the weight and
it is `imaginary' in its character, so that elements from $\sqrt{3} B_I(w)$ are `real'.
To get rid of $\sqrt{3}$ in our results, we can turn to rescaled imaginary elements via
\be
\tilde{G}_I(a_1,\ldots,a_{w})=\sqrt{3}\;G_I(a_1,\ldots,a_{w})\;.
\label{Gtilde}
\ee
The numbers of elements in these bases $B(w)$ are ${3, 8, 21, 55, 144}$ for weights $w=1,2,3,4,5$, correspondingly.
If a constant is expected to be uniformly transcendental one can use these bases.
Otherwise, one uses 
\bea
\bar{B}(w)=\bigcup_{i=1}^w B(i)\;.
\label{basisW}
\eea
The numbers of elements in these bases are {4, 12, 33, 88, 232} for weights $w=1,2,3,4,5$, correspondingly.

In simple situations, the number of available digits per constant in a basis can be as small as 7.
In more complicated situations, with cumbersome coefficients in results, it can be more than 15.
In our case, the accuracy of 2000 digits was quite enough to obtain results with {\tt PSLQ}
in an $\ep$-expansion up to the finite part in $\ep$, or, in other words, up to weight 4, in a straightforward way.
Still at weight 5, it looks like the coefficients in results are more cumbersome and
it is better to simplify our approach.

Let us look for uniformly transcendental threshold integrals. An analysis of results
for the corresponding on-shell integrals, i.e. at $p^2=m^2$ shows that
the integrals
\begin{equation}
\{J_4=F_{1,2,2,2,2,0,\ldots,0}, \, J_5=F_{2,2,2,2,1,0,\ldots,0}\} \,.
\label{MIs-th-ut}
\end{equation}  
are uniformly transcendental. Let us assume that these integrals at $p^2=9 m^2$ also have this property.
To check this hypothesis, we run {\tt PSLQ} on coefficients of $\ep$-expansions of these integrals,
with the use of uniformly transcendental bases $B(w)$ and arrive at the following results up to the finite part in $\ep$:
\bea
J_4 &=& 
\frac{1}{\epsilon }\Bigg(
-\frac{20}{9} \tilde{G}_I(r_2) \tilde{G}_I(0,r_2)-\frac{26}{9}
   G_R(0,0,1)\Bigg)
\nonumber\\&&\mbox{} \hspace*{-14mm}   
-\frac{16}{3} \tilde{G}_I(r_2) G_R(r_4)
   \tilde{G}_I(0,r_2)+\frac{124}{3} \tilde{G}_I(r_2)
   \tilde{G}_I(0,1,r_4)
\nonumber\\&&\mbox{} \hspace*{-14mm}      
+24 \tilde{G}_I(r_2)
   \tilde{G}_I(0,r_2,-1)-\frac{100}{9} \tilde{G}_I(0,r_2)^2+8
   G_R(0,0,r_4,1)+\frac{1153}{135}  \tilde{G}_I(r_2)^4
+O(\ep) 
  \;,
\label{resJ4}   
\nn
\eea
and 
\bea
J_5 &=& 
\frac{\tilde{G}_I(r_2)}{18 \epsilon ^3} 
 +\frac{1}{\epsilon^2}\Bigg(
\frac{5}{9} \tilde{G}_I(0,r_2)-\frac{5}{9} \tilde{G}_I(r_2)
   G_R(r_4)-G_R(-1) \tilde{G}_I(r_2)
   \Bigg)   
\nonumber\\&&\mbox{} \hspace*{-5mm}   
+\frac{1}{\epsilon}\Bigg(
-\frac{52}{9} G_R(r_4) \tilde{G}_I(0,r_2)-10 G_R(-1)
   \tilde{G}_I(0,r_2)+\frac{40}{27} \tilde{G}_I(r_2)
   \tilde{G}_I(0,r_2)+6 \tilde{G}_I(0,r_2,-1)
 \nonumber\\&&\mbox{} \hspace*{-5mm}    
   +\frac{26}{3}
   \tilde{G}_I(0,1,r_4) 
+\frac{52}{27} G_R(0,0,1)+\frac{25}{9}
   \tilde{G}_I(r_2) G_R(r_4)^2+10 G_R(-1)
   \tilde{G}_I(r_2) G_R(r_4)
\nonumber\\&&\mbox{} \hspace*{-5mm}    
   +9 G_R(-1)^2
   \tilde{G}_I(r_2)+\frac{253 \tilde{G}_I(r_2)^3}{108}
\Bigg)      
\nonumber\\&&\mbox{} \hspace*{-5mm}    
+\frac{32}{9} \tilde{G}_I(r_2) G_R(r_4)
   \tilde{G}_I(0,r_2)+\frac{1060}{27} G_R(r_4)^2
   \tilde{G}_I(0,r_2)-60 G_R(r_4)
   \tilde{G}_I(0,r_2,-1)
\nonumber\\&&\mbox{} \hspace*{-5mm}    
+104 G_R(-1) G_R(r_4)
   \tilde{G}_I(0,r_2)+\frac{5101}{324} G_R(0,0,1)
   \tilde{G}_I(r_2)+90 G_R(-1)^2 \tilde{G}_I(0,r_2)
\nonumber\\&&\mbox{} \hspace*{-5mm}    
   -54
   G_R(-1) \tilde{G}_I(0,r_2,-1)+14 \tilde{G}_I(0,r_2)
   G_R(r_2,-1)  
-96 G_R(-1)
   \tilde{G}_I(0,1,r_4)
\nonumber\\&&\mbox{} \hspace*{-5mm}      
   -\frac{530}{9} G_R(r_4)
   \tilde{G}_I(0,1,r_4)-60
   \tilde{G}_I(0,1,r_2,r_3)-\frac{248}{9} \tilde{G}_I(r_2)
   \tilde{G}_I(0,1,r_4)
\nonumber\\&&\mbox{} \hspace*{-5mm}       
+\frac{5695}{108} \tilde{G}_I(r_2)^2
   \tilde{G}_I(0,r_2)-16 \tilde{G}_I(r_2)
   \tilde{G}_I(0,r_2,-1)+\frac{200}{27}
   \tilde{G}_I(0,r_2)^2     
\nonumber\\&&\mbox{} \hspace*{-5mm}    
-\frac{7438}{81} \tilde{G}_I(0,0,0,r_2)-74
   \tilde{G}_I(0,1,r_2,-1)+54
   \tilde{G}_I(0,r_2,1,-1)  
\nonumber\\&&\mbox{} \hspace*{-5mm}    
+\frac{250}{9}
   \tilde{G}_I(0,1,1,r_4)-\frac{16}{3}
   G_R(0,0,r_4,1)-\frac{1021}{27} \tilde{G}_I(r_2)^3
   G_R(r_4)-\frac{250}{27} \tilde{G}_I(r_2)
   G_R(r_4)^3
\nonumber\\&&\mbox{} \hspace*{-5mm}    
-50 G_R(-1) \tilde{G}_I(r_2)
   G_R(r_4)^2-90 G_R(-1)^2 \tilde{G}_I(r_2)
   G_R(r_4)-\frac{287}{6} G_R(-1)
   \tilde{G}_I(r_2)^3
\nonumber\\&&\mbox{} \hspace*{-5mm}      
   -54 G_R(-1)^3
   \tilde{G}_I(r_2)-\frac{2306}{405} \tilde{G}_I(r_2)^4         
   +O(\ep)
   \;.
\label{resJ5} 
\nn
\eea
 
To evaluate the $\ep$-term of $J_1$ let us construct the following
linear combination:
\bea
J_6=\left(1+\frac{1}{2}\epsilon+\frac{95}{12}\epsilon ^2+\frac{2615 }{144}\epsilon^3+\frac{1154333 }{1728}\epsilon ^4\right) J_1 
+48 \epsilon J_4  -3024 \epsilon^3  J_5
\;.
\label{resJ6}
\eea
The coefficients here are adjusted in such a way that the available result up to the finite part in $\ep$ is
uniformly transcendental. Moreover, analytical result for its $\ep$-term can be revealed
with the help of the basis 
\bea
\tilde{B}(5) = B(5)\cup 
\left\{1, \tilde{G}_I(r_2),-\frac{20}{9} \tilde{G}_I(r_2)
   \tilde{G}_I(0,r_2)-\frac{26}{9} G_R(0,0,1)\right\}  
\;.
\nn
\eea  
which differs from the uniformly transcendental basis of weight 5 given by Eq.~(\ref{basisW}) by adding three elements
that are proportional to the leading terms of $J_1,J_5,J_4$ in their $\ep$-expansions.

Running {\tt PSLQ} on a high-precision numerical value of the $\ep$-term of $J_6$, with this basis we obtain an analytical result from which
we derive the $\ep$-term of $J_1$ and thereby arrive at the following
result for the first master integral at threshold
\bea
J_1 &=& 
-\frac{1}{4 \epsilon ^4}+\frac{1}{8 \epsilon^3} 
+\frac{1}{\epsilon ^2}\Bigg(
\frac{23}{12}-\frac{\tilde{G}_I(r_2)^2}{4}
\Bigg)
+\frac{1}{\epsilon}\Bigg(
-\frac{1}{3}
   G_R(0,0,1)+\frac{\tilde{G}_I(r_2)^2}{8}+\frac{1493}{576}
\Bigg) 
\nonumber\\&&\mbox{} 
-40 \tilde{G}_I(r_2) G_R(r_4) \tilde{G}_I(0,r_2)+60
   \tilde{G}_I(r_2) \tilde{G}_I(0,1,r_4)+\frac{320}{3}
   \tilde{G}_I(r_2) \tilde{G}_I(0,r_2)
\nonumber\\&&\mbox{}   
+72
   G_R(0,0,r_4,1) +\frac{833}{6} G_R(0,0,1)+\frac{647
   \tilde{G}_I(r_2)^4}{60}+\frac{23 \tilde{G}_I(r_2)^2}{12}
\nonumber\\&&\mbox{}   
   +168
   \tilde{G}_I(r_2)+\frac{1024805}{6912}  
\nonumber\\&&\mbox{}   
+ \epsilon  \Bigg(
-352 \tilde{G}_I(r_2) G_R(r_4)^2
   \tilde{G}_I(0,r_2)-864 G_R(-1) \tilde{G}_I(r_2)
   \tilde{G}_I(0,1,r_4)
\nonumber\\&&\mbox{}  
   +276 \tilde{G}_I(r_2) G_R(r_4)
   \tilde{G}_I(0,r_2)
+528 \tilde{G}_I(r_2) G_R(r_4)
   \tilde{G}_I(0,1,r_4)
  \nonumber\\&&\mbox{} 
   +864 \tilde{G}_I(r_2) G_R(r_4)
   \tilde{G}_I(0,r_2,-1)-\frac{15563}{27} G_R(0,0,1)
   \tilde{G}_I(r_2)^2
 \nonumber\\&&\mbox{} \hspace*{-0mm}   
   +576 \tilde{G}_I(r_2) \tilde{G}_I(0,r_2)
   G_R(r_2,-1)  
+864 \tilde{G}_I(r_2)
   \tilde{G}_I(0,1,r_2,r_3)-2014 \tilde{G}_I(r_2)
   \tilde{G}_I(0,1,r_4)
\nonumber\\&&\mbox{} \hspace*{-0mm}    
   -960 \tilde{G}_I(r_2)
   \tilde{G}_I(0,1,1,r_4)+568 \tilde{G}_I(0,r_2)
   \tilde{G}_I(0,1,r_4)   
-\frac{72172}{81} \tilde{G}_I(r_2)^3
   \tilde{G}_I(0,r_2)
\nonumber\\&&\mbox{} \hspace*{-0mm}    
   +\frac{320}{27} \tilde{G}_I(r_2)
   \tilde{G}_I(0,r_2)-1152 \tilde{G}_I(r_2)
   \tilde{G}_I(0,r_2,-1)   
+\frac{14816}{9} \tilde{G}_I(r_2)
   \tilde{G}_I(0,0,0,r_2)
\nonumber\\&&\mbox{} \hspace*{-0mm}      
   +288 \tilde{G}_I(r_2)
   \tilde{G}_I(0,1,r_2,-1)+\frac{1600}{3}
   \tilde{G}_I(0,r_2)^2+1680 \tilde{G}_I(0,r_2)
\nonumber\\&&\mbox{} \hspace*{-0mm}    
   +1136
   G_R(0,0,1,r_2,r_4) 
 +288 G_R(r_4)
   G_R(0,0,r_4,1)-420 G_R(0,0,r_4,1)
\nonumber\\&&\mbox{} \hspace*{-0mm}     
   -288
   G_R(0,0,1,1,r_4)+\frac{485}{27}
   G_R(0,0,1)-\frac{397811}{405} G_R(0,0,0,0,1)
 \nonumber\\&&\mbox{} \hspace*{-0mm}   
   +\frac{5132}{15}
   \tilde{G}_I(r_2)^4 G_R(r_4)
-1680 \tilde{G}_I(r_2)
   G_R(r_4)+168 G_R(-1) \tilde{G}_I(r_2)^4
\nonumber\\&&\mbox{} \hspace*{-0mm}   
   -3024
   G_R(-1) \tilde{G}_I(r_2)-\frac{29905
   \tilde{G}_I(r_2)^4}{72}+\frac{1493
   \tilde{G}_I(r_2)^2}{576}
\nonumber\\&&\mbox{} \hspace*{-0mm}   
+\frac{27244
   \tilde{G}_I(r_2)}{9}+\frac{232538063}{82944}
\Bigg)                
    +O(\ep^2)  \;.
\label{resJ1}   
\nn
\eea

We apply this procedure based on uniformly transcendental bases also to $J_2$ and $J_3$.
Here we use, in a similar way, the following two linear combinations
\bea
J_7&=& 
\left(1+\frac{1}{3}\epsilon +\frac{37}{9} \epsilon ^2+\frac{571 }{108}\epsilon^3+\frac{139585 }{324}\epsilon ^4\right) J_2 
-37 \epsilon J_4  +2112 \epsilon^3 J_5 
\;, 
\label{resJ7} \\
J_8&=& 
\left(1+8 \epsilon ^2-\frac{277 }{2}\epsilon ^3-\frac{29551 }{12}\epsilon ^4\right)J_3 
+8  (6 \epsilon -1) J_4 +16 (743 \epsilon +48) \epsilon ^2 J_5
\;.
\label{resJ8}
\eea
Then for the second and the third master integrals at threshold, we obtain the following results up to $\ep^1$:
\bea
J_2 &=& 
-\frac{1}{4 \epsilon ^4}
+\frac{1}{\epsilon^2}\Bigg( 2-\frac{\tilde{G}_I(r_2)^2}{4}\Bigg)
\nonumber\\&&\mbox{} \hspace*{-0mm}  
+\frac{1}{\epsilon}\Bigg(
-\frac{160}{9} \tilde{G}_I(r_2) \tilde{G}_I(0,r_2)-\frac{211}{9}
   G_R(0,0,1)-\frac{128 \tilde{G}_I(r_2)}{3}-\frac{277}{8}
\Bigg)
\nonumber\\&&\mbox{} \hspace*{-0mm}  
+ 40 \tilde{G}_I(r_2) G_R(r_4) \tilde{G}_I(0,r_2)-60
   \tilde{G}_I(r_2) \tilde{G}_I(0,1,r_4)-\frac{740}{9}
   \tilde{G}_I(r_2) \tilde{G}_I(0,r_2)
\nonumber\\&&\mbox{} \hspace*{-0mm}   
 -72
   G_R(0,0,r_4,1)-107 G_R(0,0,1)-\frac{647}{60}
   \tilde{G}_I(r_2)^4-\tilde{G}_I(r_2)^2-\frac{352
   \tilde{G}_I(r_2)}{3}-\frac{1647}{16}    
\nonumber\\&&\mbox{} \hspace*{-0mm}   
+ \epsilon  \Bigg(   
352 \tilde{G}_I(r_2) G_R(r_4)^2 \tilde{G}_I(0,r_2)+864
   G_R(-1) \tilde{G}_I(r_2)
   \tilde{G}_I(0,1,r_4)
\nonumber\\&&\mbox{} \hspace*{-0mm}    
   -\frac{632}{3} \tilde{G}_I(r_2)
   G_R(r_4) \tilde{G}_I(0,r_2)
-528 \tilde{G}_I(r_2)
   G_R(r_4) \tilde{G}_I(0,1,r_4)
\nonumber\\&&\mbox{} \hspace*{-0mm}     
   -864 \tilde{G}_I(r_2)
   G_R(r_4) \tilde{G}_I(0,r_2,-1)+\frac{15563}{27}
   G_R(0,0,1) \tilde{G}_I(r_2)^2
 \nonumber\\&&\mbox{} \hspace*{-0mm}     
   -576 \tilde{G}_I(r_2)
   \tilde{G}_I(0,r_2) G_R(r_2,-1) 
-864 \tilde{G}_I(r_2)
   \tilde{G}_I(0,1,r_2,r_3)
\nonumber\\&&\mbox{} \hspace*{-0mm}     
   +\frac{4648}{3} \tilde{G}_I(r_2)
   \tilde{G}_I(0,1,r_4)+960 \tilde{G}_I(r_2)
   \tilde{G}_I(0,1,1,r_4)-568 \tilde{G}_I(0,r_2)
   \tilde{G}_I(0,1,r_4) 
\nonumber\\&&\mbox{} \hspace*{-0mm}     
+\frac{72172}{81} \tilde{G}_I(r_2)^3
   \tilde{G}_I(0,r_2)+\frac{3310}{81} \tilde{G}_I(r_2)
   \tilde{G}_I(0,r_2)+888 \tilde{G}_I(r_2)
   \tilde{G}_I(0,r_2,-1)
\nonumber\\&&\mbox{} \hspace*{-0mm}    
-\frac{14816}{9} \tilde{G}_I(r_2)
   \tilde{G}_I(0,0,0,r_2)-288 \tilde{G}_I(r_2)
   \tilde{G}_I(0,1,r_2,-1)-\frac{3700}{9}
   \tilde{G}_I(0,r_2)^2
\nonumber\\&&\mbox{} \hspace*{-0mm}    
   -\frac{3520}{3} \tilde{G}_I(0,r_2)   
-1136
   G_R(0,0,1,r_2,r_4)-288 G_R(r_4)
   G_R(0,0,r_4,1)
\nonumber\\&&\mbox{} \hspace*{-0mm}    
   +320 G_R(0,0,r_4,1)+288
   G_R(0,0,1,1,r_4)+\frac{4195}{81}
   G_R(0,0,1)
\nonumber\\&&\mbox{} \hspace*{-0mm}    
   +\frac{397811}{405} G_R(0,0,0,0,1)  
-\frac{5132}{15}
   \tilde{G}_I(r_2)^4 G_R(r_4)+\frac{3520}{3}
   \tilde{G}_I(r_2) G_R(r_4)
 \nonumber\\&&\mbox{} \hspace*{-0mm}    
   -168 G_R(-1)
   \tilde{G}_I(r_2)^4+2112 G_R(-1)
   \tilde{G}_I(r_2)   
+\frac{34517 \tilde{G}_I(r_2)^4}{108}
\nonumber\\&&\mbox{} \hspace*{-0mm} 
-\frac{31
   \tilde{G}_I(r_2)^2}{48}-\frac{53122
   \tilde{G}_I(r_2)}{27}-\frac{1046989}{576}
\Bigg)                 
   +O(\ep^2)
\;,
\label{resJ2}  
\nn
\eea
\bea
J_3 &=& 
-\frac{1}{4 \epsilon ^4}
+\frac{1}{\epsilon^2}\Bigg( 2-\frac{\tilde{G}_I(r_2)^2}{4}\Bigg)
\nonumber\\&&\mbox{}
+\frac{1}{\epsilon}\Bigg(
-\frac{160}{9} \tilde{G}_I(r_2) \tilde{G}_I(0,r_2)-\frac{211}{9}
   G_R(0,0,1)-\frac{128 \tilde{G}_I(r_2)}{3}-\frac{277}{8}
\Bigg)
\nonumber\\&&\mbox{} \hspace*{-0mm}    
-\frac{248}{3} \tilde{G}_I(r_2) G_R(r_4)
   \tilde{G}_I(0,r_2)+\frac{1172}{3} \tilde{G}_I(r_2)
   \tilde{G}_I(0,1,r_4)   
+\frac{320}{3} \tilde{G}_I(r_2)
   \tilde{G}_I(0,r_2)
 \nonumber\\&&\mbox{} \hspace*{-0mm}   
   +192 \tilde{G}_I(r_2)
   \tilde{G}_I(0,r_2,-1)-\frac{800}{9}
   \tilde{G}_I(0,r_2)^2   
-\frac{1280}{3} \tilde{G}_I(0,r_2)+136
   G_R(0,0,r_4,1)
\nonumber\\&&\mbox{} \hspace*{-0mm}     
   +\frac{416}{3} G_R(0,0,1)+\frac{1280}{3}
   \tilde{G}_I(r_2) G_R(r_4)  
+768 G_R(-1)
   \tilde{G}_I(r_2)+\frac{42719 \tilde{G}_I(r_2)^4}{540}
\nonumber\\&&\mbox{} \hspace*{-0mm}     
   +2
   \tilde{G}_I(r_2)^2-\frac{5944
   \tilde{G}_I(r_2)}{9}-\frac{30319}{48}    
\nonumber\\&&\mbox{} \hspace*{-0mm}    
+ \epsilon  \Bigg(
-\frac{25952}{27} \tilde{G}_I(r_2) G_R(r_4)^2
   \tilde{G}_I(0,r_2)-3552 G_R(-1) \tilde{G}_I(r_2)
   \tilde{G}_I(0,1,r_4)
\nonumber\\&&\mbox{} \hspace*{-0mm}       
   +256 \tilde{G}_I(r_2) G_R(r_4)
   \tilde{G}_I(0,r_2)  
+\frac{976}{9} \tilde{G}_I(r_2)
   G_R(r_4) \tilde{G}_I(0,1,r_4)
\nonumber\\&&\mbox{} \hspace*{-0mm}   
   +1632 \tilde{G}_I(r_2)
   G_R(r_4) \tilde{G}_I(0,r_2,-1)+\frac{16640}{9}
   G_R(r_4) \tilde{G}_I(0,r_2)^2   
+\frac{13312}{3}
   G_R(r_4) \tilde{G}_I(0,r_2)
\nonumber\\&&\mbox{} \hspace*{-0mm}    
   +\frac{53165}{243}
   G_R(0,0,1) \tilde{G}_I(r_2)^2-1728 G_R(-1)
   \tilde{G}_I(r_2) \tilde{G}_I(0,r_2,-1)
\nonumber\\&&\mbox{} \hspace*{-0mm}    
   +640
   \tilde{G}_I(r_2) \tilde{G}_I(0,r_2)
   G_R(r_2,-1)  
+7680 G_R(-1) \tilde{G}_I(0,r_2)+3552
   \tilde{G}_I(r_2) \tilde{G}_I(0,1,r_2,r_3)
\nonumber\\&&\mbox{} \hspace*{-0mm}   
   -1984
   \tilde{G}_I(r_2) \tilde{G}_I(0,1,r_4)-\frac{24320}{9}
   \tilde{G}_I(r_2) \tilde{G}_I(0,1,1,r_4)   
+\frac{34616}{9}
   \tilde{G}_I(0,r_2) \tilde{G}_I(0,1,r_4)
\nonumber\\&&\mbox{} \hspace*{-0mm}     
   -\frac{2392484}{729}
   \tilde{G}_I(r_2)^3 \tilde{G}_I(0,r_2)+\frac{1120}{9}
   \tilde{G}_I(r_2) \tilde{G}_I(0,r_2)-1152 \tilde{G}_I(r_2)
   \tilde{G}_I(0,r_2,-1)
\nonumber\\&&\mbox{} \hspace*{-0mm}     
+\frac{146656}{27} \tilde{G}_I(r_2)
   \tilde{G}_I(0,0,0,r_2)+2912 \tilde{G}_I(r_2)
   \tilde{G}_I(0,1,r_2,-1)
\nonumber\\&&\mbox{} \hspace*{-0mm}     
   -1728 \tilde{G}_I(r_2)
   \tilde{G}_I(0,r_2,1,-1)+\frac{1600}{3}
   \tilde{G}_I(0,r_2)^2   
-\frac{59440}{9} \tilde{G}_I(0,r_2)
\nonumber\\&&\mbox{} \hspace*{-0mm} 
+1920
   \tilde{G}_I(0,r_2) \tilde{G}_I(0,r_2,-1)-4608
   \tilde{G}_I(0,r_2,-1)-6656
   \tilde{G}_I(0,1,r_4)
\nonumber\\&&\mbox{} \hspace*{-0mm}    
   +\frac{119152}{9}
   G_R(0,0,1,r_2,r_4)    
-7680 G_R(0,0,1)
   G_R(r_2,-1)
\nonumber\\&&\mbox{} \hspace*{-0mm}    
   +11520 G_R(-1)
   G_R(0,0,r_2,-1)     
   +11520 G_R(0,0,1,r_2,-1)+11520
   G_R(0,0,r_2,1,-1)
\nonumber\\&&\mbox{} \hspace*{-0mm}    
   +544 G_R(r_4)
   G_R(0,0,r_4,1)-384 G_R(0,0,r_4,1)   
-544
   G_R(0,0,1,1,r_4)
\nonumber\\&&\mbox{} \hspace*{-0mm}     
   -8960 G_R(-1)^2
   G_R(0,0,1)+\frac{1480}{9} G_R(0,0,1)-\frac{33658939
   G_R(0,0,0,0,1)}{3645}
\nonumber\\&&\mbox{} \hspace*{-0mm}     
   -10240
   G_R(0,0,1,1,-1)   
+\frac{250204}{135} \tilde{G}_I(r_2)^4
   G_R(r_4)-\frac{6400}{3} \tilde{G}_I(r_2)
   G_R(r_4)^2
\nonumber\\&&\mbox{} \hspace*{-0mm}       
   -7680 G_R(-1) \tilde{G}_I(r_2)
   G_R(r_4)
+\frac{59440}{9} \tilde{G}_I(r_2)
   G_R(r_4)-\frac{1480}{3} G_R(-1)
   \tilde{G}_I(r_2)^4
\nonumber\\&&\mbox{} \hspace*{-0mm}      
   -6912 G_R(-1)^2 \tilde{G}_I(r_2)+11888
   G_R(-1) \tilde{G}_I(r_2) 
-\frac{18448
   \tilde{G}_I(r_2)^4}{45}-\frac{16192
   \tilde{G}_I(r_2)^3}{9}
\nonumber\\&&\mbox{} \hspace*{-0mm}       
   -\frac{277
   \tilde{G}_I(r_2)^2}{8}  
-\frac{170156
   \tilde{G}_I(r_2)}{27}-\frac{1763005}{288}
\Bigg)                              
 +O(\ep^2)
\;.
\label{resJ3}   
\nn
\eea

\section{Conclusion}
 
Using our algorithm to solve differential equations by expansions near singular points we obtained
high-precision values for our master integrals at threshold and then arrived, with the use of the {\tt PSLQ} algorithm, 
at analytical values. In other words, with our procedure we have transported simple information about 
the master integrals in the large-momentum limit to the complicated point $p^2=9 m^2$ and obtained there analytical results.
Moreover, starting from our boundary conditions at infinity, we analyzed not only the `naive' part of 
threshold expansion but also leading terms of the form $(9-x)^{n-6\ep}$ ($n$ is integer) and observed that the 
same bases also work and lead to analytical results via the {\tt PSLQ} algorithm.
Besides, proceeding in a similar way we arrived at the (Taylor) expansions at the singular point
$x=0$, with coefficients in terms of elements of our bases, in agreement with results for 
vacuum integrals \cite{Czakon:2004bu,Schroder:2005va}.
Therefore, we have demonstrated that although a canonical form of differential equations in our example 
is impossible and we don't know analytical results for the integrals, we can obtain analytical results 
for these integrals at some special kinematic points where the integrals are expressed in terms of usual polylogarithmic constants.
 
We have obtained results up to $\ep^1$ but we believe that multiple polylogarithms values at sixth roots of unity 
form bases also at higher weights. The only possible complication when going to higher orders is  connected with 
the fact that the size of the bases rapidly grows with the transcendental weight. We would like to emphasize that 
the bottleneck here is not connected with obtaining high-precision results using \texttt{DESS}, but rather 
with subsequent applications of the {\tt PSLQ} algorithm.
In fact, after the current calculation was done, we realized that we might use smaller (by 20-25 percents) bases
defined in Ref.~\cite{Kniehl:2017ikj} via values of harmonic polylogarithms at sixth roots of unity.
At least, the results presented in this paper can be expressed also in terms of these constants.
Of course, it should be simpler to try to extend these results to higher weights using these bases.

\vspace{0.2 cm}
{\em Acknowledgments.}
V.S. is grateful to Michail Kalmykov and Oleg Veretin for instructive discussions.
The work of A.S. and V.S. was supported by RFBR, grant 17-02-00175A.
The work of R.L. was supported by the grant of the `Basis' foundation for theoretical physics.

\end{document}